\documentclass{IOS-Book-Article}

\usepackage{mathptmx}
\usepackage{booktabs} 
\usepackage{xspace}
\usepackage{tikz}
\usetikzlibrary{patterns,calc}
\usepackage{pgf-umlsd}
\usepackage{algorithm}
\usepackage{algorithmic}
\usepackage{pifont}
\usepackage{hyperref}
\usepackage{float}
\usepackage{wrapfig}
\restylefloat{figure}

\def\hb{\hbox to 10.7 cm{}}

\newcommand*\blackcircled[1]{\tikz[baseline=(char.base)]{
    \node[shape=circle,fill,inner sep=0.5pt] (char) {\textcolor{white}{#1}};}}
\newcommand{\projName}{\textsc{Mqt-Tz}\xspace}
\newcommand{\sys}{\projName}
\newcommand{\arm}{{Arm}\xspace}
\newcommand{\optee}{\textsc{Op-Tee}\xspace}
\newcommand{\tz}{\textsc{TrustZone}\xspace}

\usepackage{tcolorbox}
\usepackage{fancyhdr}
\begin{document}

\pagestyle{fancy}

\begin{frontmatter}              

\title{\projName: Secure MQTT Broker for Biomedical Signal Processing on the Edge}

\markboth{}{April 2020\hb}

\author[A,B]{\fnms{Carlos} \snm{Segarra}%
\thanks{Corresponding Author: Carlos Segarra; E-mail: carlossegarragonzalez@gmail.com.}},
\author[A]{\fnms{Ricard} \snm{Delgado-Gonzalo}}
and
\author[B]{\fnms{Valerio} \snm{Schiavoni}}

\runningauthor{C. Segarra et al.}
\address[A]{CSEM, Neuch\^atel, Switzerland}
\address[B]{Universit\'e de Neuch\^atel, Neuch\^atel, Switzerland}

\begin{abstract}
Physical health records belong to healthcare providers, but the information contained within belongs to each patient. In an increasing manner, more health-related data is being acquired by wearables and other IoT devices following the ever-increasing trend of the \textit{Quantified Self}. Even though data protection regulations (\textit{e.g.}, GDPR) encourage the usage of privacy-preserving processing techniques, most of the current IoT infrastructure was not originally conceived for such purposes. One of the most used communication protocols, MQTT, is a lightweight publish-subscribe protocol commonly used in the Edge and IoT applications. In MQTT, the broker must process data on clear text, hence exposing a large attack surface for a malicious agent to steal/tamper with this health-related data. In this paper, we introduce \sys, a secure MQTT broker leveraging \arm \tz, a popular Trusted Execution Environment (TEE). We define a mutual TLS-based handshake and a two-layer encryption for end-to-end security using the TEE as a trusted proxy. We provide quantitative evaluation of our open-source PoC on streaming ECGs in real time and highlight the trade-offs.
\end{abstract}

\end{frontmatter}
\markboth{September 2019\hb}{September 2019\hb}

\section{Introduction} \label{sec:introduction}
Personalized health and medicine has the potential of being the next revolution in healthcare. It is also referred as the P4 medicine (Predictive, Preventive, Personalized, and Participatory), and provides the opportunity to benefit from more targeted and effective diagnoses and treatments~\cite{Cumming2014}. One of the driving forces behind this tendency is the increasing medicalization of wearable technology~\cite{Dunn2018} and mobile health (mHealth) apps~\cite{Gagnon2015}. In order to enable these technologies, complex processing IoT pipelines are gradually being deployed or repurposed. When the data-in-motion are vital signs, protecting user's privacy becomes a topic of crucial importance. Recent data protection regulations (\textit{e.g.}, GDPR~\cite{UniEuropeanParlimentCouncilEuropean2016}) stress the importance of protecting sensitive information against malicious attackers or untrusted cloud providers.

Message Queuing Telemetry Transport (MQTT)~\cite{Banks2014} is one of the most commonly-used communication protocols in IoT. In spite of that, it is not included in some of the most extended  Medical-Grade data exchange standards~\cite{Continua2018,IHE2015}. It follows a publish-subscribe architecture specially designed for environments with limited memory and reduced network bandwidth. In such circumstances, MQTT has proven to be more adapted to the IoT than classical protocols such as HTTP~\cite{Yokotani2016}. In MQTT, a \textit{client} publishes data to a \textit{topic} and the \textit{broker} forwards it to each client that previously subscribed to it. The protocol is currently used in a variety of settings: data generation by sensors, pre-processing on the edge, and forwarding to the cloud. Examples include live heart-rate data~\cite{Chooruang2016,Segarra2019}, smart-grids~\cite{Krylovskiy2015}, or building management systems~\cite{Lee2016}. Most MQTT implementations support TLS for transport security in the client-broker link, preventing malicious actors from spoofing application data. However, the broker itself still exposes a great attack surface~\cite{TeseraktAG2019}.

In order to protect the privacy of health-related data, we present \projName, a secure implementation of the MQTT broker leveraging \arm \tz, a Trusted Execution Environments (TEE), widely available on edge devices~\cite{Liu2018}. \tz is a security feature available in recent \arm processors that enables system-wide hardware isolation for trusted software~\cite{Amacher2019}. Our prototype builds atop \texttt{mosquitto} (\url{https://mosquitto.org}), a popular MQTT broker implementation, and includes persistent storage of client's keys in \arm's tamper-proof secure storage, as well as TEE-protected re-encryption of application data. These security enhancements make our ecosystem compliant with the \textit{"Services Secure Interface"}~\cite{Continua2018} described by the Personal Connected Health Alliance, and address several attack vectors listed~\cite{IHE2015} by the IHE. We also consider linking our secure broker to a larger storage utility where data-at-rest is encrypted and its origin authenticated by \sys.

The paper is organized as follows. In Section~\ref{sec:mqttz}, we describe the technical architecture and implementation of \projName. Then, in Section~\ref{sec:evaluation}, we evaluate its performance and robustness at processing 1-lead ECGs in real time. Finally, in Section~\ref{sec:conclusion} we expose our main conclusions and propose further lines of research.
\begin{wrapfigure}{R}{.55\linewidth}
    \resizebox{\linewidth}{!}{
        \input{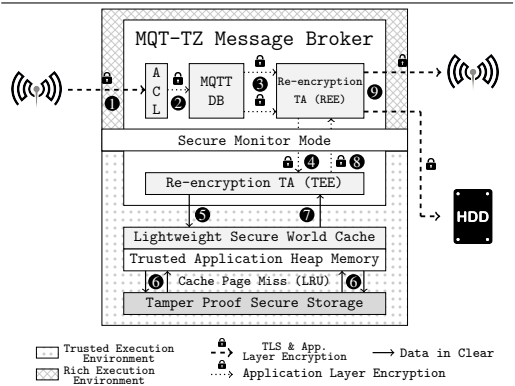}
    }
    \caption{\projName Architecture and data flow.\label{fig:architecture}}
\end{wrapfigure}

\section{\sys: Securing the MQTT Broker}\label{sec:mqttz}
\subsection{Architecture \& Component Description}\label{sec:architecture}
\tz splits the system in a hardware-protected trusted part (the TEE) and an untrusted one (also called Rich Execution Environment, or REE). We add an encryption layer in MQTT's payload using client-specific keys stored in \arm's secure storage. This way, application data is only processed in clear inside the TEE. For the additional key-provisioning, we redefine the client authentication in the mutual TLS handshake to prevent the REE from gaining access to clients' keys.

The overall workflow looks as follows. Data travels two-fold encrypted from the client to the broker (Fig.\ref{fig:architecture}-\ding{182}). Once the client access is confirmed, Fig.\ref{fig:architecture}-\ding{183}, the subscribers for the given topic are retrieved and the payload forwarded (Fig.\ref{fig:architecture}-\ding{184}). Then, encrypted data is transferred to the TEE (Fig.\ref{fig:architecture}-\ding{185}). The origin and destination client keys are retrieved (\ding{186}-\ding{188}), the payload is re-encrypted, and sent back to the REE (Fig.\ref{fig:architecture}-\ding{189}) and to the subscriber (Fig.\ref{fig:architecture}-\ding{190}).

\textbf{Two-Step Handshake.}
\projName defines and uses a two-step handshake that realizes broker and client authentication with end-to-end encryption from the client to the TEE.  The handshake protocol requires minimal pre-provisioned cryptographic material. The broker (server in TLS nomenclature) authentication is done through TLS' handshake, supported by default in \texttt{mosquitto}. The client authentication is done through MQTT. It publishes its symmetric key, encrypted with the broker's TEE public key, to a specific write-only topic. This TEE key-pair is generated at device start-up time (secure boot) and derived from a Hardware Unique Key (HUK).

\textbf{Layered Encryption \& Access Control Mechanisms.}
Once the initial handshake is finished, \projName uses a two-layer encryption mechanism. First, the client-broker link is protected by TLS within MQTT. Second, MQTT's payload field is encrypted using the clients' symmetric key. Then, data is re-encrypted in the TEE (explained next) and sent to destination over MQTT-TLS. Doing so, we achieve end-to-end security relying on \tz as a secure proxy.

\textbf{Payload Re-encryption.}
The core secure functionality implemented in \projName is the payload re-encryption. We link MQTT with a Trusted Application (TA) running inside the TEE that transfers the encrypted data to the Secure World, retrieves the origin and destination keys from secure storage, and re-encrypts the information. Currently, topic subscription lists and MQTT metadata are stored in a dedicated database (MQTT DB) in the REE. We plan on shadowing these structures and keeping them in the TEE.

\textbf{Lightweight Cache.}
\sys embeds a lightweight cache that keeps the most recent keys in the TA's heap memory, and evicts the least used to persistent secure storage.

\subsection{Implementation Details} \label{sec:implementation}
\projName is implemented in \texttt{C}. The current version of \projName adds 400 SLOC to \texttt{mosquitto} and the TA amounts to 1184 SLOC.  The \sys TA relies on \optee (\url{https://www.optee.org}), an open-source framework with native support for \tz. Our implementation will be publicly available (\url{https://github.com/mqttz}).

\textbf{Client and Server Authentication.}
The server-side authentication is done through vanilla TLS. We deploy \projName's secure broker in a device with a static IP address.  Then, we bound the address to a domain name and use a certificate. We rely on \textit{Let's Encrypt} (\url{https://letsencrypt.org/}) to get one and to authenticate the broker. The client-side authentication uses MQTT as communication layer, and \texttt{openssl} (v1.1.1a) for cryptographic primitives and operations. The integration with \texttt{mosquitto} exploits custom callbacks for each packet processing. In addition, we use MQTT Request/Response (RR) features (since v5) for the client's key exchange. To control access and R/W permissions to topics, we use \texttt{mosquitto}'s ACLs.

\textbf{Trusted Application.}
We use \optee to implement the payload re-encryption TA. Code developed within this framework has two parts: \emph{(1)}, a host app that runs in the REE and acts as entry point and bridge to the TEE, and \emph{(2)} a trusted API in the TEE that exposes different functions. \projName intercepts all MQTT packets being forwarded to the recipients, and feeds our host app with both client's IDs, and the encrypted data. We then perform the payload re-encryption using \optee's storage and cryptographic libraries. \tz not only provides isolation between worlds, but also between different TAs. Hence, we use the same secure API to store new keys during the handshake. For the key retrieval, we plan to implement a small LRU cache to store the most frequently used keys in the TA's heap, and the rest in persistent secure storage.

\section{Evaluation and Results} \label{sec:evaluation}
In this section, we perform an evaluation of \projName. First, we benchmark the TA re-encryption with random data in order to understand the overhead introduced by the re-encryption; and then, we analyze the CPU and network throughput when monitoring vital signs in a real setting. For all experiments, we virtualize a Raspberry Pi 3 using \textsc{Qemu-v8} (\url{https://www.qemu.org/}) running \texttt{mosquitto} v1.6.3 and \optee v3.5.0.

\subsection{TA Re-encryption}
\begin{figure*}[t!]
    \centering
    \includegraphics[width=\linewidth]{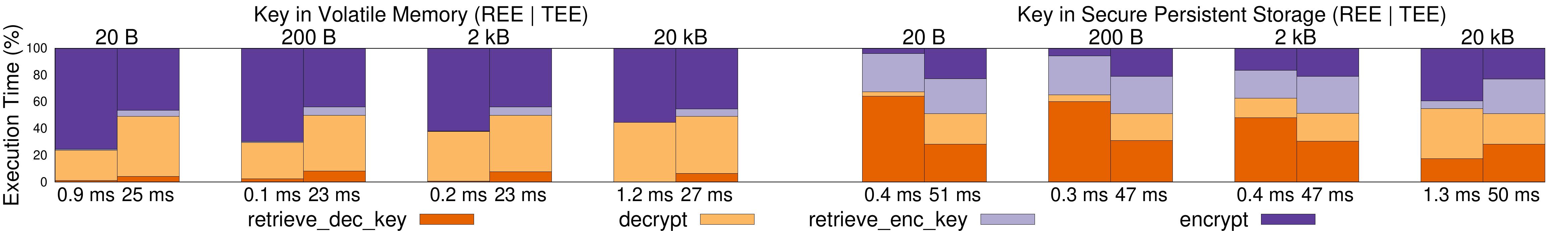}
    \vspace{-10pt}
    \caption{Re-encryption TA microbenchmark.\label{fig:ub1}}
\end{figure*}

In Fig.~\ref{fig:ub1}, we show the breakdown of the time required to re-encrypt a single block of data for different sizes. The time is split in the time to retrieve each key (\texttt{retrieve\_dec\_key}, \texttt{retriev\_enc\_key}), \texttt{encrypt}, and \texttt{decrypt}. We can observe that AES is two orders of magnitude slower in the TEE. This is a consequence of \optee not using hardware accelerators in contrast to \texttt{openssl} in the REE. Moreover, we observe sensible slowdowns when switching from in-memory to secure persistent storage.

\subsection{Real-time ECG Processing}
\begin{wrapfigure}{R}{.51\linewidth}
    \resizebox{\linewidth}{!}{
        \includegraphics[width=\linewidth]{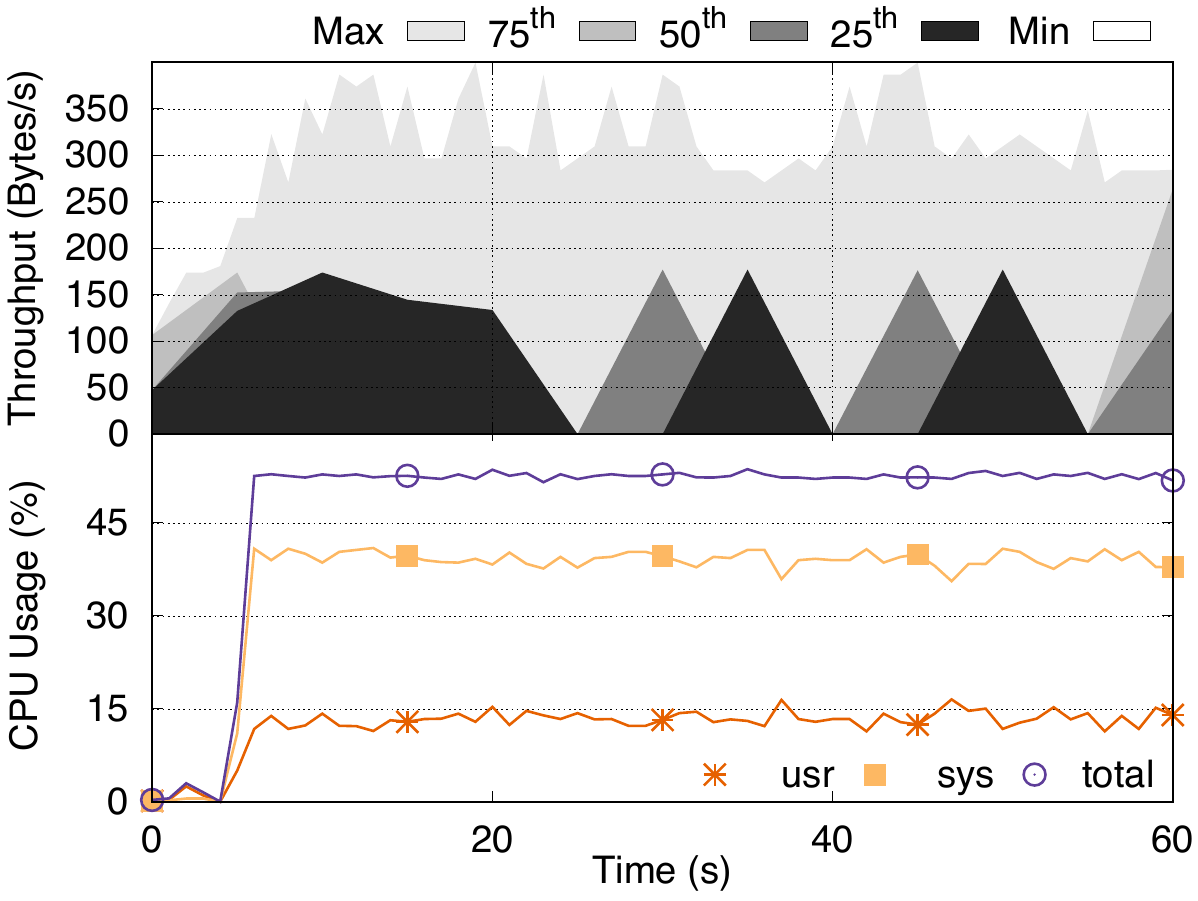}
    }
    \vspace{-10pt}
    \caption{Workload test: Network throughput (top) and CPU usage (bottom)\label{fig:ecg-test}}
\end{wrapfigure}

In this case, we test the resilience of \sys at sustaining the workloads that can be encountered in a hospital. For the experiments, we use the LTMS-S~\cite{Chetelat2015} platform developed by CSEM for the European Space Agency (ESA). In particular, we simulate 50 patients streaming in real-time 1-lead electrocardiograms (ECGs) at a frequency of 321.25~Hz. All ECGs are streamed toward a single \sys broker. In Fig.~\ref{fig:ecg-test}, we depict the outbound throughput generated by each publisher measured using \texttt{nethogs} (\url{https://github.com/raboof/nethogs}). We observe that at any given time only a subset of the publishers actually emits data. A single subscriber streams at 350~Bytes/s in the worst case, and the full collective generates between 3 to 5~kBytes per second. During the experiment, we recorded using \texttt{dstat} (\url{https://github.com/dagwieers/dstat}) the CPU load at the broker, shown in Fig.~\ref{fig:ecg-test}. We observe that after the initial peak, the overall CPU usage (both for \texttt{usr} and \texttt{sys} processes) stabilizes at 60\%.

\section{Conclusion and Future Work} \label{sec:conclusion}
Motivated by the lack of secure-by-design communication protocols for the Edge, we presented \projName, our secure implementation of the MQTT broker using \tz and showed its direct application in a in-hospital setting. The proposed system is robust and capable of managing 50 patients in real-time with a CPU usage of 60\%. We plan to extend this work along the following directions. First, we will compare \sys against other publish-subscribe protocols and messaging queues. Second, we will study the performance overhead of \sys when deployed on large-scale scenarios. Finally, we intend to look into the energy footprint, an aspect of paramount relevance for edge deployments.
\vspace{-15pt}

\bibliographystyle{ieeetr}
\bibliography{biblio}

\begin{thebibliography}{10}

\bibitem{Cumming2014}
G.~P. Cumming, ``Connecting \& collaborating - {H}ealthcare for the 21st
  century,'' in {\em PAHI'2014}, 2014.

\bibitem{Dunn2018}
J.~Dunn, R.~Runge, and M.~Snyder, ``Wearables and the medical revolution,''
  {\em Pers. Med.}, vol.~15, no.~5, pp.~429--448, 2018.

\bibitem{Gagnon2015}
M.-P. Gagnon, P.~Ngangue, J.~Payne-Gagnon, and M.~Desmartis, ``m-{H}ealth
  adoption by healthcare professionals: {A} systematic review,'' {\em J. Am.
  Med. Inform. Assn.}, vol.~23, pp.~212--220, June 2015.

\bibitem{UniEuropeanParlimentCouncilEuropean2016}
{The European Parliment and the Council of the European Union}, ``Regulation
  ({EU}) 2016/679,'' 2016.

\bibitem{Banks2014}
A.~Banks and R.~Gupta, ``{MQTT} version 3.1.1,'' software, OASIS, Oct. 2014.

\bibitem{Continua2018}
{Personal Connected Health Alliance}, ``Fundamentals of medical-grade data
  exchange,'' white paper, Continua, Sept. 2018.

\bibitem{IHE2015}
{IHE PCD Technical Committee}, ``Medical equipment management ({MEM}):
  {M}edical device cyber security,'' white paper, IHE International, Inc., Oct.
  2015.

\bibitem{Yokotani2016}
T.~Yokotani and Y.~Sasaki, ``Comparison with {HTTP} and {MQTT} on required
  network resources for {IoT},'' in {\em ICCEREC'2016}, pp.~1--6, Sept. 2016.

\bibitem{Chooruang2016}
K.~Chooruang and P.~Mangkalakeeree, ``Wireless heart rate monitoring system
  using {MQTT},'' {\em Procedia Comput. Sci.}, vol.~86, pp.~160--163, 2016.

\bibitem{Segarra2019}
C.~Segarra, R.~Delgado-Gonzalo, M.~Lemay, P.-L. Aublin, P.~Pietzuch, and
  V.~Schiavoni, ``Using trusted execution environments for secure stream
  processing of medical data,'' in {\em Lect. Notes Comput. Sc.}, vol.~11534,
  pp.~91--107, 2019.

\bibitem{Krylovskiy2015}
A.~Krylovskiy, M.~Jahn, and E.~Patti, ``Designing a smart city internet of
  things platform with microservice architecture,'' in {\em FiCloud'2015},
  pp.~25--30, Aug. 2015.

\bibitem{Lee2016}
Y.~Lee, H.~Hsiao, C.~Huang, and S.~T. Chou, ``An integrated cloud-based smart
  home management system with community hierarchy,'' {\em IEEE. T Consum.
  Electr.}, vol.~62, pp.~1--9, Feb. 2016.

\bibitem{TeseraktAG2019}
{Teserakt AG}, ``Is {MQTT} secure? (a report),'' 2019.

\bibitem{Liu2018}
R.~Liu and M.~Srivastava, ``{VirtSense}: {V}irtualize sensing through {ARM}
  {TrustZone} on {Internet-of-Things},'' in {\em SysTEX'2018}, (New York, NY,
  USA), pp.~2--7, ACM, 2018.

\bibitem{Amacher2019}
J.~Amacher and V.~Schiavoni, ``On the performance of {ARM} {TrustZone},'' in
  {\em DAIS'2019}, pp.~133--151, 2019.

\bibitem{Chetelat2015}
O.~Ch{\'e}telat, D.~Ferrario, M.~Proen{\c{c}}a, J.-A. Porchet, A.~Falhi,
  O.~Grossenbacher, R.~Delgado-Gonzalo, N.~Della~Ricca, and C.~Sartori,
  ``Clinical validation of {LTMS-S}: {A} wearable system for vital signs
  monitoring,'' in {\em EMBC'2015}, pp.~3125--3128, 2015.

\end{thebibliography}
  
\end{document}